\documentclass[conference]{IEEEtran}
\IEEEoverridecommandlockouts

\usepackage{cite}
\usepackage{amsmath,amssymb,amsfonts}
\usepackage{algorithmic}
\usepackage{graphicx}
\usepackage{textcomp}
\usepackage{xcolor}

\usepackage{multirow}
\usepackage{makecell}
\usepackage{amsmath}
\usepackage[capitalise]{cleveref}

\usepackage{flushend}

\usepackage{physics}

\def\BibTeX{{\rm B\kern-.05em{\sc i\kern-.025em b}\kern-.08em
    T\kern-.1667em\lower.7ex\hbox{E}\kern-.125emX}}

\usepackage{comment}
\excludecomment{confidential}

\begin{document}

\title{DisQu: Investigating the Impact of Disorder in Quantum Generative Models \thanks{We gratefully acknowledge financial support from the Quantum Initiative Rhineland-Palatinate QUIP and the Research Initiative Quantum Computing for AI (QC-AI). \textit{(Nikolaos Palaiodimopoulos and Maximilian Kiefer-Emmanouilidis contributed equally to this work.)} \textit{(Corresponding author: Nikolaos Palaiodimopoulos)}}}


\author{
\IEEEauthorblockN{
Yannick Werner \IEEEauthorrefmark{1},
Jasmin Frkatovic\IEEEauthorrefmark{2},
Vitor Fortes Rey\IEEEauthorrefmark{1}\IEEEauthorrefmark{2},
Matthias Tschöpe\IEEEauthorrefmark{1},
Sungho Suh\IEEEauthorrefmark{1}\IEEEauthorrefmark{2},\\
Paul Lukowicz\IEEEauthorrefmark{1}\IEEEauthorrefmark{2},
Nikolaos Palaiodimopoulos \IEEEauthorrefmark{1}\IEEEauthorrefmark{2},
and Maximilian Kiefer-Emmanouilidis\IEEEauthorrefmark{1}\IEEEauthorrefmark{2}\IEEEauthorrefmark{3}}
\IEEEauthorblockA{\IEEEauthorrefmark{1}German Research Center for Artificial Intelligence (DFKI), Kaiserslautern, Germany}
\IEEEauthorblockA{\IEEEauthorrefmark{2}Department of Computer Science and Research Initiative QC-AI, RPTU Kaiserslautern-Landau, Kaiserslautern, Germany}
\IEEEauthorblockA{\IEEEauthorrefmark{3}Department of Physics, RPTU Kaiserslautern-Landau, Kaiserslautern, Germany}
}

\maketitle

\begin{abstract}
Disordered Quantum many-body Systems (DQS) and Quantum Neural Networks (QNN) have many structural features in common. However, a DQS is essentially an initialized QNN with random weights, often leading to non-random outcomes. In this work, we emphasize the possibilities of random processes being a deceptive quantum-generating model effectively hidden in a QNN. When we choose weights in a QNN randomly the unitarity property of quantum gates is unchanged. As we show, this can lead to memory effects with multiple consequences on the learnability and trainability of QNN one would not expect from a classical neural network with random weights. This phenomenon may lead to a fundamental misunderstanding of the capabilities of common quantum generative models, where the generation of new samples is essentially averaging over random outputs. While we suggest that DQS can be effectively used for tasks like image augmentation, we draw the attention that overly simple datasets are often used to show the generative capabilities of quantum models, potentially leading to overestimation of their effectiveness.
\end{abstract}


\section{Introduction}
\label{sec:intro}

The exploration of new technological trends to solve previously hard problems is a natural progression in scientific discovery. The current speed at which quantum effects are explored and considered beneficial for artificial intelligence (AI) has led to a lot of shortcomings in the art of benchmarking and the selection of considerable datasets \cite{bowles2024better}. Following these trends could lead to the capabilities of quantum AI being completely misinterpreted. 
In this paper, we emphasize how random quantum augmentation, which introduces coherent noise into the system, can be misinterpreted as a learnable quantum generative process. We focus on quantum localization effects common in quantum-disordered many-body Systems (DQS) known to be memory-preserving beyond a threshold of randomness \cite{Abanin2019}.

Image manipulation using localization effects has already been discussed in previous works by considering quantum-disordered systems as a form of quantum-inspired augmentation \cite{Palaiodimopoulos2024} or signal decomposition and reconstruction \cite{Dutta2021, Dutta2022}. Although these contributions are a special case of quantum systems that can be understood from wave mechanics, we surprisingly found that the generic full quantum setup similarly explores memory effects under similar strengths of randomness even for amplitude-encoded complex images. 
While these quantum memory effects are expected to vanish at large system sizes and longer evolution times \cite{Suntajs2019, Suntajs2020, sels2021thermalization, sels2020dynamical, KieferUnanyan2, KieferUnanyan3, KieferUnanyan4, sierant2021observe, morningstar2021avalanches, Abanin2021}, they may prevail in artificial structures such as quantum neural networks (QNN) \cite{ Abbas2021, Tangpanitanon_2020}. Trained DQS have already been considered as an effective generative model, for example in quantum Born machines \cite{zhong2022} and under more general consideration in driven systems \cite{Tangpanitanon_2020, Tangpanitanon_2023}. However, the latter case corresponds rather to quantum random circuits \cite{Fisher23} than the DQS we investigate here.

In Ref. \cite{Palaiodimopoulos2024}, the authors present a comprehensive introduction to disordered systems and their associated localization effects, along with a clear explanation of Dirac's bra-ket notation for readers unfamiliar with the concept. For further details, we encourage readers to consult this reference. The presented paper here serves two purposes: first, to introduce DQS as a relevant technique for data augmentation, such as blurring images on quantum devices, and second, to highlight the deceptive generative capabilities of this augmentation, which may already be embedded within current QNN. 

This paper is structured as follows. In \cref{sec:method}, we introduce the augmentation protocols and similarity measures. In \cref{sec:pca}, we demonstrate how we use Principal Component Analysis (PCA) to analyze the effects of augmentation by applying it to MNIST digits \cite{lecun2010mnist}. In \cref{sec:dqgm}, we investigate which QNN structures might be most prone to disorder effects. In \cref{sec:qnns}, we present and analyze specific examples of QNNs. Our results are discussed and summarized in \cref{sec:discussion}.

\section{Quantum Augmentation}\label{sec:method}

The augmentation procedure is described by a Schrödinger time-evolution generated by a Hamiltonian of a DQS. We restrict our analysis to one-dimensional systems, where the memory effect, known as Many-Body Localization (MBL) \cite{Abanin2019}, is expected to occur at sufficiently strong disorder. Here the dynamics are presumed to be tremendously slow, and because we can control the timescales in the procedure, we can be sure that initial information is not completely lost. In the following, we relate a finite timescale of a DQS which is expected to show MBL to a single QNN layer.
The Hamiltonian has the following form,
\begin{equation}
    \label{micro_model}
    \hat H=J\sum_j(\hat{c}_j^\dagger \hat{c}_{j+1} +h.c.) +V\sum_j \hat{n}_j \hat{n}_{j+1} + \sum_j d_j \hat{n}_j,
\end{equation}
where $\hat{c}_j^\dagger (\hat{c}_j)$ are fermionic creation (annihilation) operators and $\hat{n}_j = \hat{c}_j^\dagger \hat{c}_j$ is the particle number operator acting on site $j$. Here, each lattice site can be understood as a qubit. Furthermore, $J$ is the hopping amplitude, $V$ is the nearest-neighbor interaction, and the potential disorder is drawn from a box distribution, $d_j\in [-d/2,d/2]$. Setting $V=0$ reduces the problem to single-particle effects, which are shown in \cite{Palaiodimopoulos2024}. We will use $J$ as a unit of energy and will be setting $\hbar=V=1$.
This system, and its spin-$1/2$ analogue, the XXZ model, have been studied extensively in the regime $d_c>15 J$ where MBL was believed to occur \cite{Abanin2019}. A direct implementation of the disordered XXZ model on quantum devices is presented in Ref. \cite{Smith2019}. Recent studies suggest that the disorder strength required to truly localize the system is above $d_c>80J$ \cite{sels2021markovian, morningstar2021avalanches}. It should be taken into account that this lower bound is still under debate. Thus, we should expect slow dynamics affecting the presented augmentation procedure in contrast to the non-interacting case which is proven to be Anderson localized \cite{Anderson1958} and thus a memory. 

We will use amplitude encoding to write the input images into a quantum register \cite{mottonen2004transformation, Vartiainen2004} where we follow exactly \cite{Palaiodimopoulos2024}, which, on the one hand, is one of the most generic ways to encode data on a quantum device but, on the other hand, exceeds, due to complexity, the limits of existing hardware. Since we have to work with the full quantum state to understand the augmentation effects, running the routines on quantum hardware is out of the scope of this work. New algorithms for amplitude encoding could lead to a significant boost, reducing the complexity to polynomial in time \cite{Araujo2024}. Furthermore, we present augmentation protocols that yield a possible use case on integrated quantum devices \cite{Wang2020}, for specifically tailored tasks. 

After preparing the image in a quantum register, we follow the dynamics as described by the Schrödinger equation 
$\mathrm{i}\frac{\partial}{\partial t} \ket{\psi (t)}= H\ket{\psi(t)}.$
The final state can then be computed from $\ket{\psi(t)}= \mathrm{exp}({-\mathrm{i}}Ht) \ket{\psi (t=0)},$
where one has to compute the matrix exponential \cite{HOGBEN2011179,KUPROV201131}. We set the time to $t=10\ 1/J$ commonly throughout our protocols, which is a parameter that could, in principle, be controlled in the future. Furthermore, we refer to states in the computational basis with Latin letters i.e. $\ket{j}$ and for general states with Greek letters i.e. $\ket{\psi}= \sum_{j}c_j \ket{j}$, where $c_j$ are complex values.

\begin{figure}[!t]
\centering
\includegraphics[width=0.8\columnwidth]{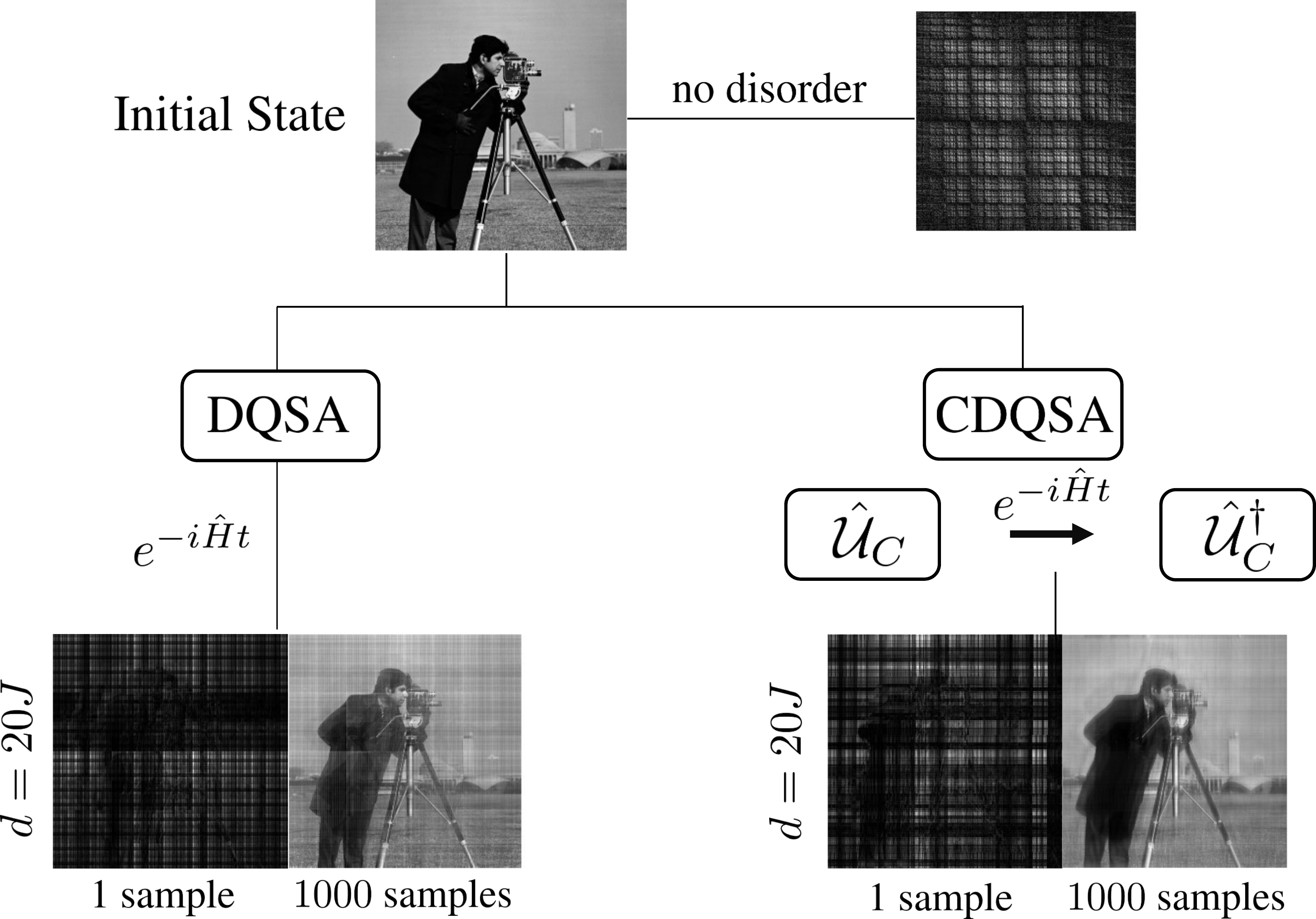}
\caption{We present the evolution of the initial image under the different quantum augmentation protocols. On the top right, we depict the evolved image without disorder, while on the bottom left and right we give indicative examples of the evolved images for the DQSA and CDQSA protocols, respectively. For each protocol, we present two images, one corresponding to a single sample and another where we have taken the average over $1000$ samples, in both cases the disorder strength is $d=20J$.}
 \label{fig:im1}
 \vspace{-5mm}
 \end{figure}

\subsection{Augmentation protocols}\label{sec:Augmentation} 
We present two different protocols for possible quantum augmentation via DQS. The protocols follow the quantum-inspired augmentation already presented in \cite{Palaiodimopoulos2024}. However, the quantum system in this contribution has no known classical simplifications, thus, one needs to consider the entire Hilbert space of dimension $2^n$, where n is the number of qubits or lattice sites. 

The first protocol will be referred to as Disordered Quantum many-body System Augmentation (DQSA) (for schematic, see Fig.~\ref{fig:im1}) and consists of the following steps:
\vspace{-2pt}
\begin{enumerate}
    \item Encode image into quantum state $\ket{\Psi_I}$.
    \item Initialize random DQS Hamiltonian $\hat H$. 
    \item Time-evolve state to time $t$ via Schrödinger equation.
    \item Save magnitudes of the final state $|\bra{j}\ket{\Psi_f}|$, reshape back to image.
    \item Repeat the previous steps $R$ times and take the average over all final states.
\end{enumerate}
The second protocol will be referred to as Cyclic Disordered Quantum many-body System Augmentation (CDQSA) (for schematic, see Fig.~\ref{fig:im1}) follows:
\vspace{-2pt}
\begin{enumerate}
    \item Initialize random Hamiltonian $\hat {H}$.
    \item Encode image into quantum state $\ket{\Psi_I}$.
    \item Make random cyclic permutation on the components of the initial state vector by applying $\hat{\mathcal{U}}_{C}$.
    \item Time-evolve state to time $t$ via Schrödinger equation.
    \item Revert random cyclic permutation by applying $\hat{\mathcal{U}}_{C}^\dagger$.
    \item Save magnitudes of the final state $|\bra{j}\ket{\Psi_I}|$, reshape back to image.
    \item Repeat steps $3-6$ $R$ times and take the average over all final states.
\end{enumerate}
Here, CDQSA can be performed on integrated quantum devices, as no disorder needs to be redrawn. Furthermore, for the interpretation of a QDS and QNN it means that a single disorder realization in combination with a suitable embedding (here cyclic permutation) is already enough to envelop memory effects.

\subsection{Similarity measures}\label{sec:similar}
We follow \cite{Palaiodimopoulos2024} to evaluate the similarity of our augmentation procedure with the original image. The first measure is
the Structural Similarity Index Metrics (SSIM) \cite{wang2004image}, which for two image windows $x$,$y$ of the equal size is defined as
\begin{equation} \label{eq:ssim}
\mathrm{SSIM}(x,y)=(l(x,y))^{\alpha}(c(x,y))^\beta(s(x,y))^\gamma.
\end{equation}
Here $l$ is the luminance, $c$ the contrast, and $s$ the correlation. The exponents are usually taken to be $\alpha=\beta=\gamma=1$.

The second measure is the Bhattacharyya coefficient\cite{fukunaga2013introduction}
\begin{align}\label{eq:magniover}
    \mathcal{B}=\sum_j^{2^n} \sqrt{P_I(j)P_f(j)}=\sum_j^{2^n} \left| \bra{j}\ket{\psi_I}\right|\left| \bra{j}\ket{\psi_f}\right|,
\end{align}
where $P_I(j)= \left| \bra{j}\ket{\psi_I}\right|^2 $ and $P_f(j)= \left| \bra{j}\ket{\psi_f}\right|^2 $. The Bhattacharyya coefficient quantifies the similarity of two probability distributions, or here the overlap of initial and final magnitude at the $j$-th computational basis state.

\begin{figure}[!t]
 \centering
 \includegraphics[width=0.8\columnwidth]{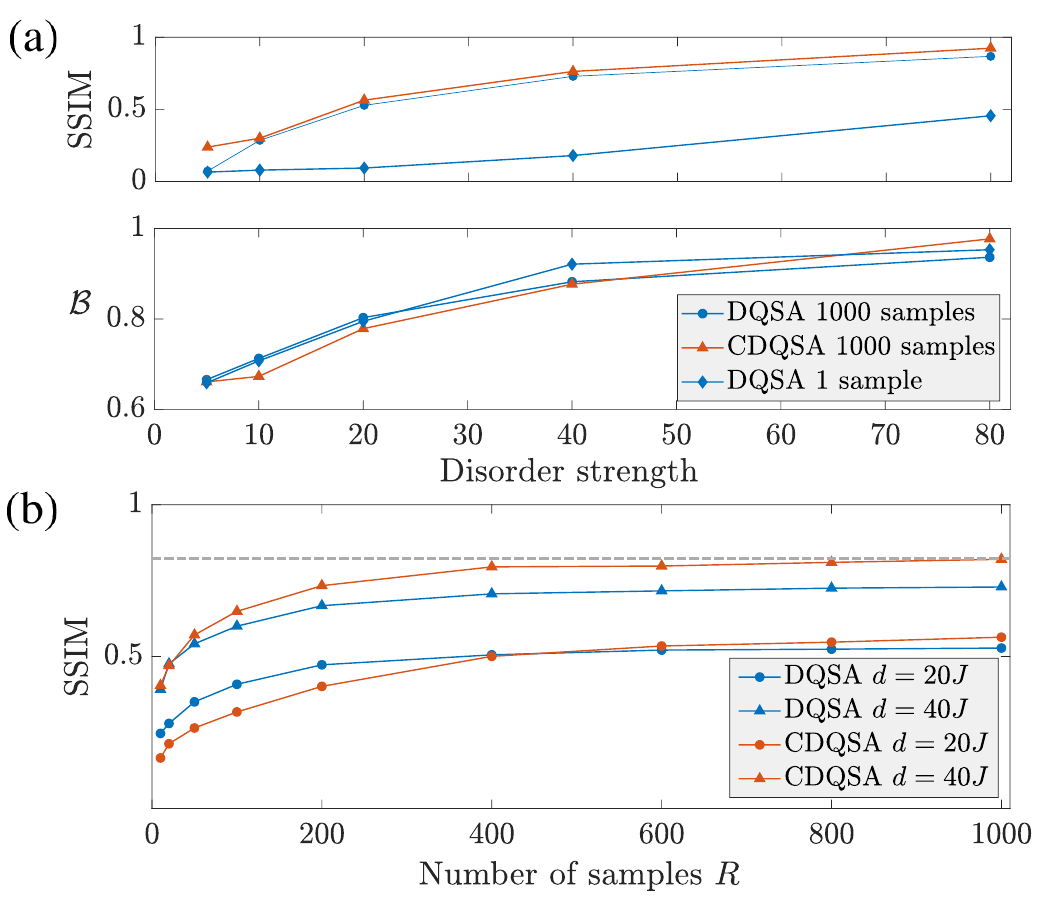}
\caption{(a) SSIM and Bhattacharyya coefficient for various disorder strengths. For both DQSA and CDQSA we plot the average over $1000$ samples, while for DQSA we also include $1$ sample. (b) SSIM as a function of the number of samples $R$ for two different disorder strengths.}
 \label{fig:im2}
 \vspace{-5mm}
 \end{figure}

The image we chose to demonstrate the effect of the quantum augmentation is a grayscale image, known as cameraman \cite{cameraman}, the size of which is $256 \times 256$ pixels. The image is encoded into the computational basis states, and we end up with a state vector of $2^n$ entries and a Hamiltonian matrix of dimensions $2^n \times 2^n$, where $n=16$. In Fig.~\ref{fig:im2}(a) we compare the behavior of the SSIM and Bhattacharyya coefficient as functions of disorder strength for the two protocols described in Sec.~\ref{sec:Augmentation}. Meanwhile, in Fig.~\ref{fig:im2}(b) we plot the SSIM as a function of the number of samples $R$ for two different disorder strengths.

The quantum augmentation technique via DQS, even though does not reach the efficiency of the quantum-inspired protocol developed in \cite{Palaiodimopoulos2024}, performs remarkably well. For both QDSA and CDQSA protocols when inspecting the retrieved images for $1$ sample (see Fig.~\ref{fig:im1}) , we can see the resulting image quality due to disorder strength alone and compare it with the one obtained after the averaging is performed. The SSIM captures the enhanced quality due to averaging (Fig.~\ref{fig:im2}(a) top panel) in contrast to the Battacharyya coefficient which is not sensitive to local properties of the state (Fig.~\ref{fig:im2}(a) bottom panel), irrespective of the particular protocol we employ. We also observe, that for moderate disorder strengths, the SSIM values for both protocols are almost stabilized after $R=400$ realizations (see Fig.~\ref{fig:im2}(b)).

\section{PCA analysis and MNIST dataset}
\label{sec:pca}

\begin{figure}[!t]
 \centering
 \includegraphics[width=\columnwidth]{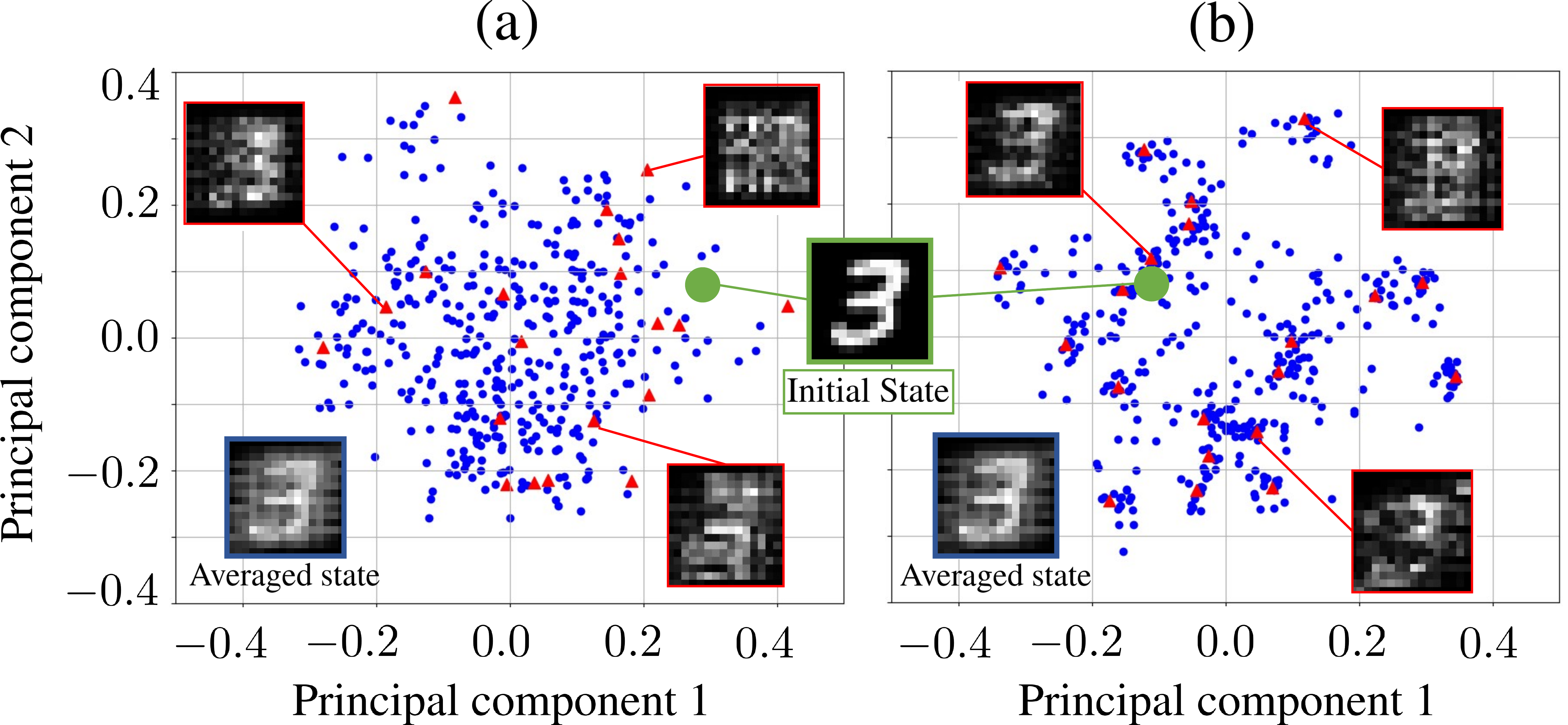}
\caption{We plot the first two principal components of the dataset, which includes the initial state (green dot), the first set of evolved states (red triangles), and the second set of evolved states (blue dots). (a) For a disorder strength of $d=10J$ and we also show the initial state (green outline), indicative examples of evolved states from the first set (red outline), and the average over the second set of evolved states (blue outline). (b) In this case, the disorder strength during the second evolution is increased to $d=20J$. In both cases $r1=20$ and $r2=20$. }
 \label{fig:im3}
 \vspace{-5mm}
 \end{figure}

In this section, we switch to the very simple MNIST handwritten digits dataset, which is widely used in the field of quantum machine learning (QML) \cite{bowles2024better}. 
To show the disorder effects, we initially selected a particular image of the digit 3, however, any MNIST image would be suitable. For convenience, the image is resized to $16\times16$ and then amplitude encoded to the quantum register. The initial state is first evolved via the DQS Hamiltonian $R_{1}$ times, each with a different disorder realization of strength $d=20J$. The resulting states make up the set of the first evolved states. Subsequently, each of the $R_{1}$ evolved states is used as an initial state which is then evolved $R_{2}$ times, with different disorder realizations of strength $d=10J$ (Fig.~\ref{fig:im3}(a)) and $d=20J$ (Fig.~\ref{fig:im3}(b)). The resulting states will be referred to as the second evolved states.

To visualize the set of first and second evolved states along with the initial state in a lower-dimensional space, we will employ PCA \cite{hotelling1933analysis}. We have reduced the dimensionality of each state to two principal components, and the results are illustrated in Fig.~\ref{fig:im3}. We want to emphasize that at no point we use PCA to reduce the dimensionality of MNIST, but only for visualization of the DQS time-evolved data. As pointed out by \cite{bowles2024better} dimensionality reduction with PCA could lead to crucially misleading results already.

We chose to use the set of first evolved states as initial states, for further evolving the system under different disorder realizations, to examine how increasing randomness would affect the retrieval of the initial image. Inspecting the 2D plot of the first two principal components (Fig.~\ref{fig:im3}), we observe that the first evolved states follow no geometric pattern and are randomly distributed around the initial state. The same holds (as expected) for the second evolved states, which form clusters only when the disorder strength during the second evolution is increased (see Fig.~\ref{fig:im3}(b)). Nevertheless, when averaging over the evolved states, we yet again retrieved a blurred image of the initial image (see insets in Fig.~\ref{fig:im3}(a-b)). This is even more surprising when one inspects some of the images of the evolved states (see images in Fig.~\ref{fig:im3}(a-b)), which hold no resemblance to the initial image. 

\section{From Augmentation to Deceptive Quantum Generative Model}
\label{sec:dqgm}

First, let us highlight the structural connections between QNN and DQS. Here we follow the general structure of QNN as described in \cite{Abbas2021}. The QNN has three distinguishable parts: encoding, trainable layers, and measurement. For our investigation, we focus on the trainable part consisting of a large unitary gate $\mathbf{U}$ with a decomposition in $k$ unitary layers $\mathbf{U}(\Vec{\theta}_k,\Vec{\theta}_{k-1,\dots,\Vec{\theta}_1})=U(\Vec{\theta}_k)U(\Vec{\theta}_{k-1})...U(\Vec{\theta}_{1})$, which are either the trainable quantum layers with weight $\Vec{\theta}$ in the QNN or represent the DQS time-evolution in a common Trotterized setup \cite{Smith2019}. However, in the latter case, $\Vec{\theta}$ are random. We emphasize that for a QNN to be considered a DQS, only the weights are selected randomly. As a result, the QNN setup when initialized by random weights resembles a disordered system. 
Still, disordered systems should not be confused with random quantum circuits \cite{Fisher23}, where unitary gates may be switched-out during the time evolution or where circuits are subjected to intermediate measurements.
 Furthermore, we constrain our research to nearest-neighbor interactions. Given the topology
 of current Near Term Noisy Quantum (NISQ) \cite{Preskill2018} devices, most QNN layers contain only gates connecting nearest-neighbor qubits. Deep QNN and multiple layers (where the layer count largely exceeds the number of qubits) will probably break this condition. In QNN, one can investigate the statistics of optimized weights, to determine if they resemble randomly drawn values. Or by checking the eigenvalue statistics of the effective Hamiltonian
\begin{equation}
H_\mathrm{eff}=\mathrm{i}\hspace{1pt}\mathrm{ln}(U(\Vec{\theta}_k)U(\Vec{\theta}_{k-1})...U(\Vec{\theta}_{1})),
\end{equation}
with methods shown in \cite{Tangpanitanon_2020}.
 
\begin{figure}[!t]
\centering
\includegraphics[width=1.0\columnwidth]{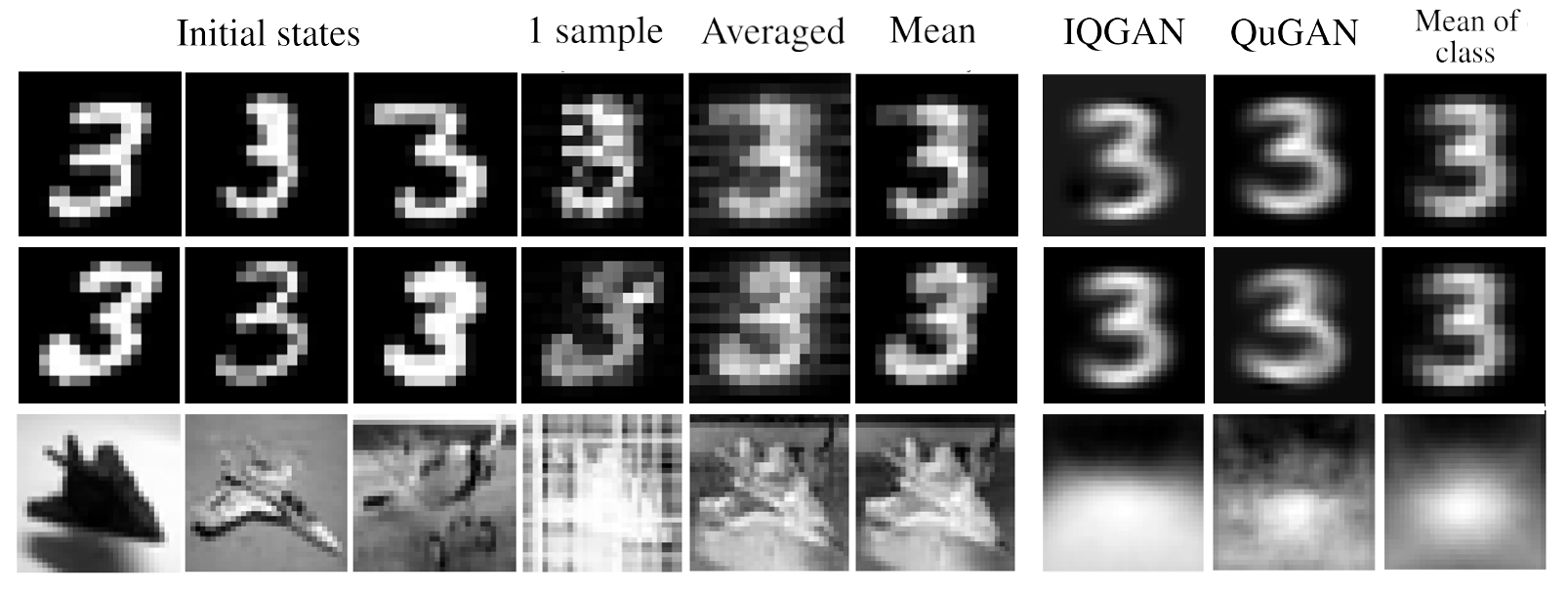}
\caption{The first three columns show the initial states for the DQSA protocol. the first two rows are selected from the MNIST dataset, while the third one is from the CIFAR-10 dataset. The next three columns display one sample of an evolved state, the average overall evolved states, and the mean of the initial set of states, respectively. On the right, we show generated images from IQGAN \cite{chu2023iqgan} and QuGAN \cite{QuGAN2021} followed by the mean of the MNIST class 3 and the CIFAR-10 class zero (plane).}  
\label{fig:im4}
\vspace{-5mm}
\end{figure}

To demonstrate how the preservation of information during the presented DQS augmentation can lead to misleading conclusions concerning the learning capabilities of generative models, we will employ the MNIST dataset. We start with three different images, of the digit 3 for MNSIT, which are our initial states (see the first two rows of Fig.~\ref{fig:im4}), and evolve each of them following the DQSA protocol (Sec.~\ref{sec:Augmentation}) across $R = 100$ disorder realizations with a strength of $d = 80J$. At the end of the evolution, we average over all the evolved states. The results, along with the initial states, are illustrated in Fig.~\ref{fig:im4}. The first two rows correspond to two different sets of initial states. Inspecting the resulting image for each case, one might mistakenly believe it is a novel generation simply because it closely resembles the initial states. However, this resemblance does not indicate that the model possesses the ability to generate truly new samples. Instead, the resulting image is merely a blurred representation of the mean of each set of initial states, capturing nothing more than the average of the original dataset. At the end of the first two rows we show the generated images from two current Quantum Generative Adversarial Networks (GAN) QuGAN \cite{QuGAN2021} and the current state-of-the-art model IQGAN \cite{chu2023iqgan}. We chose these two examples as they present a quantum formulation of a GAN where both generator and discriminator are quantum-circuits, and thus a QNN. We observe that the generated outputs for these two models closely resemble the mean of the training dataset. An effective way to test the actual generative capabilities of a model is to use a more complex dataset. This is precisely what we do in the third row of Fig.~\ref{fig:im4}, where we employ the CIFAR-10 dataset \cite{alex2009learning}. In this case, due to the greater variability in the dataset, the mean of the initial states no longer resembles a novel generated sample. In the last row, we show the images generated by QuGAN and IQGAN after being trained on CIFAR-10 class zero. As can be observed, both models fail to perform efficiently and break down.

We next examine two specific examples of QNN structures and witness how the presence or absence of localization effects affects the dynamics and to which extend the initial state is preserved.

\section{From Quantum System to QNN}\label{sec:qnns}
\begin{figure*}[!h]
    \centering
    \includegraphics[width=\textwidth]{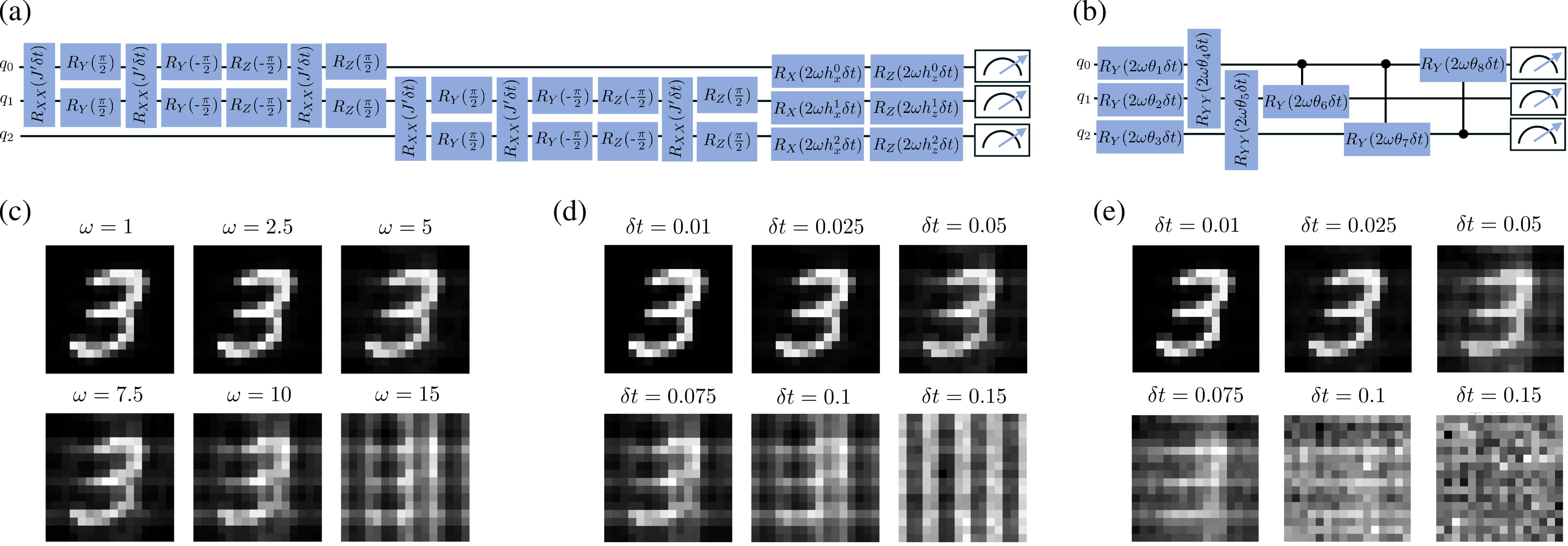}
    \caption{(a) Circuit representing a single Trotter step, as proposed in Ref. \cite{zhu2021probing}, for a system of 3 qubits. (b) QuGAN circuit from Ref. \cite{QuGAN2021}, with random weights drawn from a uniform distribution. (c) Simulated time-evolution of the initial state using the circuit shown in (a), for different scaling parameters $\omega$. Two Trotter steps are considered, each with a size of $\delta t = 0.05/J^{\prime}$. (d) Simulated time evolution of the initial state using the circuit shown in (a) for different Trotter step sizes. Here only one layer is considered and the scaling parameter is set to $\omega = 10J^{\prime}$, ensuring the system is in the MBL phase. (e) Evolved state using one layer of the QuGAN circuit, for different values of $\delta t$, with $\omega = 10J^{\prime}$. In all figures, $\omega$ is given in units of $J^{\prime}$ and $\delta t$ in units of $1/J^{\prime}$.}
    \label{fig:qnns}
\end{figure*}

As mentioned in the previous section, the architecture of the trainable layers in a QNN can be designed either to achieve a specific computational goal or to simulate the dynamics of a quantum system within a Trotterized framework. In the latter case, the presence or absence of localization effects can be known a priori, whereas in the former, the dynamics require investigation. In this section, we examine two circuits that have been used in previous works, each corresponding to one of the aforementioned cases.

The first circuit we examine was introduced in Ref. \cite{zhu2021probing} and is shown in Fig. \ref{fig:qnns}(a). This circuit employs a specific combination of gates to simulate the time evolution of the following Hamiltonian within a Trotterized framework,
\begin{equation}\label{Hamil}
\hat{H}^{\prime} = J^{\prime} \sum_i \hat{\vec{\sigma}}_i \hat{\vec{\sigma}}_{i+1} + \omega \left(\sum_i h_x^i \hat{\sigma}_x + \sum_i h_z^i \hat{\sigma}_z \right),
\end{equation}
here $J^{\prime}$ denotes the hopping amplitude, $\omega$ the scaling factor, $h_z^i$ and $h_x^i$ represent the longitudinal and transverse local fields in the $z$ and $x$ directions, respectively, and $\hat{\vec{\sigma}} = (\hat{\sigma}_x, \hat{\sigma}_y, \hat{\sigma}_z)$ is a vector operator with components corresponding to the Pauli matrices. The values of the longitudinal and transverse local fields are drawn uniformly from the interval $h_x^i, h_z^i \in [-1,1]$, indicating that we are dealing with a DQS where the scaling factor $\omega$ serves as the disorder strength. We take $h_x^i$ and $h_z^i$ to be dimensionless and the scaling factor $\omega$ is expressed in units of $J^{\prime}$. We note that, if the transverse local field was absent (i.e. $h_x^i=0$, $\forall i$) the Hamiltonian (\ref{Hamil}) would be equivalent to Eq. (\ref{micro_model}). 

Our initial state, $\ket{\Psi_{I}}$, is a resized MNIST image of the digit 3 encoded using amplitude encoding. Specifically, we use the same initial state as in Fig. \ref{fig:im3}. The evolved state is then computed as $\ket{\Psi_{f}} = \exp(-i\hat{H}^{\prime}t) \ket{\Psi_{I}}$. To apply the Trotter-Suzuki decomposition, the Hamiltonian (\ref{Hamil}) is divided into two components: one where the summation runs over even indices ($i \in \textrm{even}$), denoted as $\hat{H}_{\textrm{even}}^{\prime}$, and another where it runs over odd indices ($i \in \textrm{odd}$), denoted as $\hat{H}_{\textrm{odd}}^{\prime}$. Thus, the total Hamiltonian can be expressed as $H^{\prime} = \hat{H}_{\textrm{even}}^{\prime} + \hat{H}^{\prime}_{\textrm{odd}}$, thereby avoiding the need for calculating commutators. The evolution proceeds in time steps $\delta t$, which are chosen to be small, as the decomposition error scales with $\mathcal{O}(\delta t^2)$ as shown in the following equation,
\begin{equation}\label{eq:Trotter}
\text{e}^{i\hat{H}^{\prime}\delta t}=\text{e}^{i(\hat{H}^{\prime}_{\text{even}}+\hat{H}^{\prime}_{\text{odd}})\delta t} \approx \text{e}^{i\hat{H}^{\prime}_{\text{even}}\delta t}\text{e}^{i\hat{H}^{\prime}_{\text{even}}\delta t}+\mathcal{O}(\delta t^2).
\end{equation}
For further details, readers unfamiliar with the Trotter-Suzuki decomposition are encouraged to refer to Appendix \ref{sec:app1}. The circuit shown in Fig. \ref{fig:qnns}(a) represents a single Trotter step, with its size determined by $\delta t$ and with additional steps implemented by repeating the entire layer. The first part of the circuit, consisting of Ising-$XX$ gates sandwiched between rotation gates, represents the first term of the Hamiltonian (\ref{Hamil}), which describes the exchange interaction between neighboring qubits. Conversely, the two rotation gates, $R_{X}$ and $R_{Z}$, at the end of the circuit correspond to the transverse and longitudinal local fields, as described by the last two terms of Eq. (\ref{Hamil}). 


The analysis in Ref. \cite{zhu2021probing} indicates that when $\omega=10J^{\prime}$ the system is in the MBL phase. The initial state is evolved across $R = 1000$ different disorder realizations. For each realization, the output probability distribution is computed by measuring the qubits at the end of the circuit. The final output state is then obtained by averaging all measured probability distributions and reshaping the result into a grayscale image. 

In Fig. \ref{fig:qnns}(c) we show the final output image for different values of the scaling parameter $\omega$. We observe that the initial picture is not well preserved for scaling parameter values above $\omega = 5 J^{\prime}$. At first, this may seem counterintuitive, as increasing the scaling parameter effectively enhances the strength of the disorder. However, this behavior is solely due to the Trotter error. Since the leading error term of the Trotter-Suzuki decomposition is of order $\mathcal{O}(\delta t^2)$, we expect a breaking point where this error term becomes dominant, preventing the system from retaining the localized phase, even though the other parameters are set to ensure MBL. Therefore the simulated system is no longer able to retain information and the initial image is lost. To isolate this effect without accounting for the interplay between the first and second term of Hamiltonian (\ref{Hamil}), we set $\omega = 10 J^{\prime}$ (ensuring the system is in the MBL phase) and evolve the initial state across $R = 1000$ different disorder realizations for varying Trotter step sizes $\delta t$. The results are shown in Fig. \ref{fig:qnns}(d), where we observe that the breakdown becomes prominent as early as $\delta t = 0.075/J^{\prime}$.

The second circuit we examine is the generator circuit from the QuGAN model \cite{QuGAN2021}, which is shown in Fig. \ref{fig:qnns}(b). Every layer consists of a composition of single qubit unitaries given by $R_Y$-Gates, dual qubit unitaries posing as nearest-neighbor interaction via the Ising YY-interaction $R_{YY}$ and finally entanglement unitaries via controlled rotation gates. Each gate is parameterized by a random variable $\theta_i$ drawn from a uniform distribution, $\theta_i \in [-1,1]$, and scaled by $\omega$ and $\delta t$ to ensure consistency with the analysis of the previous circuit. To analyze the evolution of the Trotter error, we evolve the initial picture via a single layer of the QuGAN circuit using different time steps $\delta t$. The parameters we chose are the same as we used in the previous circuit. As observed in Fig. \ref{fig:qnns}(e) and similarly to the previous circuit, after some time step size $\delta t \sim 0.075/J^{\prime}$ we are unable to retrieve information about the initial image. This can be interpreted as follows: sufficiently strong randomness—in our case, $\delta t$—in a parametrized quantum circuit can disrupt information preservation due to localization effects. 

Finally, for both circuits, we quantify the preservation of the initial state using the two similarity measures introduced in Sec. \ref{sec:similar}. After encoding and evolving the image, we compare the averaged probabilities to the initial image using the Bhattacharyya coefficient (see Eq. (\ref{eq:magniover})) and the SSIM (see Eq. (\ref{eq:ssim})).
\begin{figure}[!h]
    \centering
    \includegraphics[width=0.9\columnwidth]{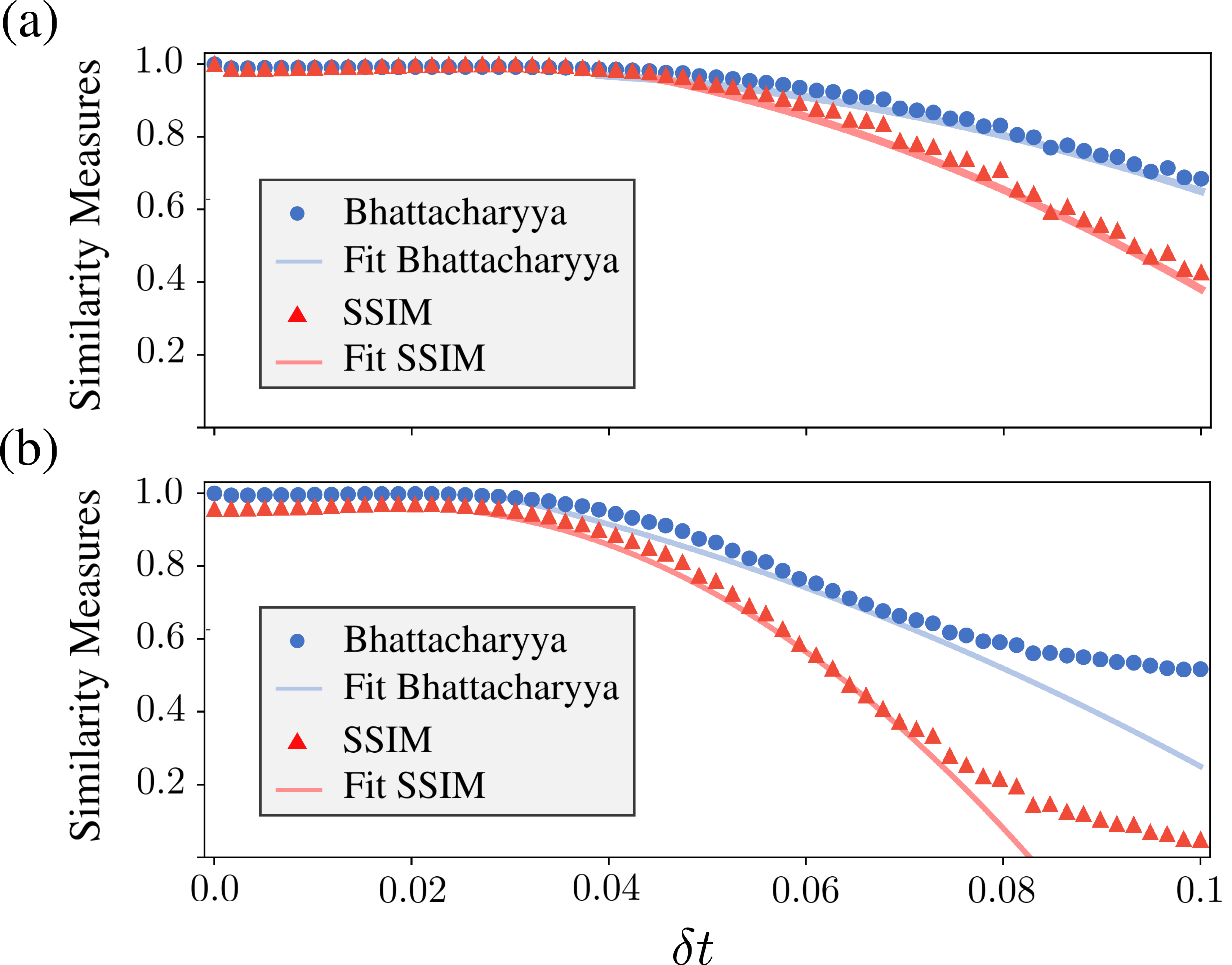}
    \caption{Similarity measures as a function of the Trotter step size. (a) The circuit under consideration is the one depicted in Fig. \ref{fig:qnns}(a). (b) Results correspond to the QuGAN circuit shown in Fig. \ref{fig:qnns}(b). For both cases, we consider a single layer with $\omega = 10J^{\prime}$. Each data point is obtained by averaging over $R = 1000$ realizations. The blue dots represent the Bhattacharyya coefficient, while the red triangles correspond to the SSIM. The faded blue and red solid lines represent the numerical fits based on Eq. (\ref{eq:sc}).}  
    \label{fig:SSIM_QGAN}
\end{figure}
By leveraging the definitions of these similarity measures alongside the Suzuki-Trotter decomposition (\ref{eq:Trotter}), we can estimate how each measure scales with the Trotter step size $\delta t$, enabling us to determine the maximum error. For convenience, we introduce a general similarity measure, $Q$, which represents both the Bhattacharyya coefficient and the SSIM, i.e., $Q \in {\mathcal{B}, S}$. We find that both quantities scale as,
\begin{equation}\label{eq:sc}
    Q \sim 1-k_{Q}\delta t^2
\end{equation}
where $k_{Q}$ is a parameter depending on the commutator of even and odd Hamiltonian parts. For a more detailed discussion, we refer the reader to Appendix \ref{sec:app2}.


In Fig. \ref{fig:SSIM_QGAN}, we plot the two similarity measures as a function of the Trotter step size for the two circuits introduced in this section. The observed behavior, consistently seen across all plots, is as follows: for very small $\delta t$, the system is not in the MBL phase, leading to a slight deterioration in image quality. As $\delta t$ increases, localization effects strengthen, slightly improving the image quality until a critical point is reached. Beyond this point, the Trotter error becomes dominant, causing the quality to degrade. Additionally, in the same plots, we compare the expected scaling of the similarity measures according to Eq. (\ref{eq:sc}). The values of the parameter $k_{Q}$ are determined numerically by optimizing the fit curve in the region where the Trotter error dominates.


\section{SUMMARY \& DISCUSSION}
\label{sec:discussion} 

In this work, we have demonstrated that the quantum-inspired augmentation protocols presented in \cite{Palaiodimopoulos2024} can be generalized and operate efficiently when the underlying dynamics are governed by a fully quantum system, in contrast to a classical interpretation using light. Remarkably, the images can be retrieved even when they are amplitude-embedded in a $2^n$-dimensional Hilbert space. The effects of augmentation on the 'cameraman' image can be understood as a form of blurring, particularly in the CDQSA case, although not without artifacts. A trained setup and fine-tuning of the time scales could help optimize the level of blurring, and this entire process could be implemented on a quantum device. The ability to recover the initial image by averaging the magnitudes of the evolved states in the presented augmentation protocols directly stems from the preservation of information during the evolution of a quantum dynamical system (QDS). This became even more evident in the PCA analysis we conducted in Sec.~\ref{sec:pca}. Furthermore, in Sec.~\ref{sec:dqgm} after establishing the connection between DQS and QNN we investigated the implications of our results to the efficiency of quantum generative models. We showed that, when we average over a set of initial states (see Fig.~\ref{fig:im4}), the corresponding generated image always resembles the mean of the considered dataset, though slightly blurred. The generating process can thus be reduced to a weighted mean procedure and has no generative capabilities. Further evidence supporting our claim were found in current Quantum GAN systems IQGAN and QuGAN, where the generated output from the trained models closely resembles the mean of the MNIST dataset. By employing the more complex CIFAR-10 dataset we demonstrated that these models fail to perform efficiently. At this point, it is necessary to mention that the usage of PCA in both models seems to overfit strongly, which is also suggested in \cite{bowles2024better}. Thus, larger parts of the circuits can be chosen randomly, without changing the results significantly. A more detailed analysis without the use of PCA is thus necessary for evaluating the proposed quantum circuits, which is, however, out of the scope of this work. Finally, in Sec.~\ref{sec:qnns}, we analyzed two QNN structures recently proposed in the literature \cite{zhu2021probing,QuGAN2021}, where the localization properties of the system result in the preservation of the initial state. By highlighting the connection between a Hamiltonian and a QNN derived from its Trotterization, we demonstrated that increasing the Trotter step size eventually causes the Trotter error to dominate, leading to the breakdown of information preservation.  

Our findings have significant implications for the design of QNNs, as they reveal fundamental properties that influence their behavior and performance. Quantum circuits that exhibit the localization properties of a DQS can be utilized to implement the augmentation protocols presented herein. However, these same properties must be approached with caution when employing such circuits for generative tasks, as they may yield deceptive results. Moreover, the perspective provided by the Trotterized framework enables us to explain why and when these properties breakdown, allowing for the informed design of QNNs. Our results pave the way for an intriguing research direction: the systematic categorization of QNN architectures—such as types of gates, number of layers, and other structural elements—based on their localization properties and their ability to preserve information.

\appendices
\section{Trotter-Suzuki decomposition} \label{sec:app1} 
Following the Trotter-Suzuki decomposition the evolution operator when considering a Hamiltonian $H=\sum_{i=1}^{n}H_{i}$ can be approximated by:
\begin{equation}
e^{-iHt} \approx \bigg( \prod_{i=1}^{n} e^{-iH_{i}\frac{t}{m}} \bigg)^m,
\end{equation}
where $m$ is the number of Trotter steps and $\delta t= t/m$. To arrive at the desired decomposition, we analyze how to write the sum of exponential functions by considering two non-commuting operators $\hat{A}$ and $\hat{B}$. To do so, we expand the exponential terms using a Taylor series expansion, introducing a small parameter $x\ll1$. Thus, we get,
\begin{equation}\label{eq:ap1}
\begin{split}
    \text{e}^{x(A+B)} =& 1 + x(A+B) + \frac{x^2}{2} (A+B)^2 + \mathcal{O}(x^3)\\
    =& 1+  x(A+B) + \frac{x^2}{2} (A^2 + 2AB + B^2) \color{black}\\ 
     &+ \frac{x^2}{2} \comm{A}{B}+ \mathcal{O}(x^3)
\end{split}
\end{equation}
and
\begin{equation}\label{eq:ap2}
\begin{split}
    \text{e}^{xA}\text{e}^{xB} =& \left(1 + xA + \frac{x^2}{2} A^2\right) \left(1+ xB + \frac{x^2}{2} B^2\right)\\
     &+  \mathcal{O}(x^3) \\
    =& 1+ x(A+B) + \frac{x^2}{2} (A^2 + 2AB + B^2) \color{black}+\mathcal{O}(x^3),
\end{split}
\end{equation}
where for simplicity the operator hat notation has been omitted.
In the first equation we used the commutator relation to change the order of the operators, i.e. $BA = AB - \comm{A}{B}$. We observe that Eq. (\ref{eq:ap1}) and Eq. (\ref{eq:ap2}), differ only in the addition of the commutator term. Therefore, omitting the higher order terms we can write,
\begin{equation}\label{eq:std}
\begin{split}
    \text{e}^{x(A+B)} &\approx \text{e}^{xA}\text{e}^{xB} + \frac{x^2}{2} \comm{A}{B} \\
    &= \text{e}^{xA}\text{e}^{xB} + \mathcal{O}(x^2),
\end{split}
\end{equation}
yielding the desired representation for the Suzuki-Trotter decomposition.

\section{Scaling of the similarity measures}
\label{sec:app2} 
To extract the scaling properties of our measures, we first reduce the problem to two unitary gates acting on two qubits separately. Furthermore, they have an overlap on the qubit in the middle. These gates correspond to the previous Trotter-Suzuki decomposition, where one unitary is created by $\hat{H}_{\text{even}}$ and one by $\hat{H}_{\text{odd}}$.  According to Eq. (\ref{eq:std}), we can express the evolved state as,
\begin{equation}\label{eq:std1}
\begin{split}
    \langle j|\Psi_f \rangle &= \langle j | e^{-i \hat{H}\delta t}|\Psi_I\rangle \\
    &=\langle j| e^{-i \hat{H}_{\text{even}}\delta t}e^{-i \hat{H}_{\text{odd}}\delta t}|\Psi_I\rangle\\
    &-\frac{\delta t^2}{2}\langle j|[\hat{H}_{\text{even}},\hat{H}_{\text{odd}}]|\Psi_I\rangle,
\end{split}
\end{equation}
where $\ket{j}$ represents a basis vector of the state space. Utilizing Eq. (\ref{eq:std1}) the Bhattacharyya coefficient can be written as:
\begin{equation}
\begin{split}
    B &= \sum_j^{2^n} |\langle j|\Psi_I\rangle||\langle j|\Psi_f\rangle|\\
    &=\sum_j^{2^n} |\langle j|\Psi_I\rangle| \left( |\langle j|\Psi_I\rangle| - \frac{\delta t^2}{2} |\langle j|[\hat{H}_{\text{even}},\hat{H}_{\text{odd}}]|\Psi_I\rangle|\right)\\
    &= \sum_j^{2^n}|\langle j|\Psi_I\rangle|^2 - \frac{\delta t^2}{2}\sum_j^{2^n}|\langle j|\Psi_I \rangle||\langle j|[\hat{H}_{\text{even}},\hat{H}_{\text{odd}}]|\Psi_I\rangle|\\
    &= 1 -\frac{\delta t^2}{2}\sum_j^{2^n} |\langle j|\Psi_I \rangle| |\langle j|[\hat{H}_{\text{even}},\hat{H}_{\text{odd}}]|\Psi_I\rangle|\\
    &=: 1 - k_{\mathcal{B}} \delta t^2,
\end{split}
\end{equation}
where we have made the following assumption: that due to strong localization by MBL and short-time dynamics, the magnitudes of the initial state is mostly conserved and therefore $|\langle j| e^{-i \hat{H}_{\text{even}}\delta t}e^{-i \hat{H}_{\text{odd}}\delta t}|\Psi_I\rangle|\approx|\langle j|\Psi_I\rangle|$.

To evaluate the scaling of the SSIM which has the following form, 
\begin{equation*}
    \textrm{SSIM} (x,y) =l(x,y)c(x,y) s(x,y),
\end{equation*}
where $l(x,y)$ and $c(x,y)$ depend only on the mean and variance of the x and y values. We assume that  $y_{i} \to x_{i}+\delta x_{i}$, where the $\delta x_{i}$ is drawn from a uniform distribution with zero mean and a small range of $\delta x$. In the limit of large sampling (standard error of the mean) the mean and variance average to their true values, i.e. $\mu_y \rightarrow \mu_x$ and $\sigma_y \rightarrow \sigma_x$. Consequently, in this simplified picture, we get $\bar{l}(x,y)=\bar{c}(x,y)=1$, where the bar indicates the sampling procedure. Thus, the scaling depends solely on $\bar{s}(x,y)$, and can be formulated as  
\begin{equation}
\text{SSIM}(x,y)\sim \frac{E[(x-\mu_x)(y-\mu_y)]}{\sigma_x\sigma_y} \sim 1 - k_S\delta x^2.
\end{equation}
\\ \\

\bibliographystyle{IEEEbib}

\end{document}